\documentclass[fullpage,10pt,doublecolumn]{article}

\usepackage{array}
\usepackage{amsmath, amsthm, amssymb} 
\usepackage{algpseudocode}
\usepackage{algorithm}
\usepackage{multirow}
\usepackage{bigstrut}
\usepackage{lastpage}

  \usepackage[dvips]{graphicx}
  \graphicspath{{./figures/}}
  \DeclareGraphicsExtensions{.eps}

\newcommand{\graphG}{\mathcal{G}}
\newcommand{\setE}{\mathbf{E}}
\newcommand{\setV}{\mathbf{V}}

\newcommand{\setR}{\mathbf{R}}
\newcommand{\setC}{\mathbf{C}}
\newcommand{\setF}{\mathbf{F}}

\newcommand{\pref}{\psi}
\newcommand{\npref}{\bar{\psi}}
\newcommand{\gain}{\gamma}
\newcommand{\rep}{\phi}
\newcommand{\ch}{\chi}

\def \figureheightfour {2.38cm}

\usepackage{color}
\newcommand{\BeginRevision}{\color{black}}%revise of the text
\newcommand{\EndRevision}{\color{black}}%comment of the text

\begin{document}

	\title{Dispersing Instant Social Video Service\\Across Multiple Clouds}
	
	\author{
			Zhi Wang\footnote{Z.~Wang is with the Graduate School at Shenzhen, Tsinghua University, Email: wangzhi@sz.tsinghua.edu.cn},
			Baochun Li,
			Lifeng Sun,
			Wenwu Zhu,
			and Shiqiang Yang
	}
	\date{}
	\maketitle

\begin{abstract} 

Instant social video sharing which combines the online social network and user-generated short video streaming services, has become popular in today's Internet. Cloud-based hosting of such instant social video contents has become a norm to serve the increasing users with user-generated contents. A fundamental problem of cloud-based social video sharing service is that users are located globally, who cannot be served with good service quality with a single cloud provider. In this paper, we investigate the feasibility of dispersing instant social video contents to multiple cloud providers. The challenge is that inter-cloud social \emph{propagation} is indispensable with such multi-cloud social video hosting, yet such inter-cloud traffic incurs substantial operational cost. We analyze and formulate the multi-cloud hosting of an instant social video system as an optimization problem. We conduct large-scale measurement studies to show the characteristics of instant social video deployment, and demonstrate the trade-off between satisfying users with their ideal cloud providers, and reducing the inter-cloud data propagation. Our measurement insights of the social propagation allow us to propose a heuristic algorithm with acceptable complexity to solve the optimization problem, by partitioning a propagation-weighted social graph in two phases: a preference-aware initial cloud provider selection and a propagation-aware re-hosting. Our simulation experiments driven by real-world social network traces show the superiority of our design.

\end{abstract}
\section{Introduction}
\label{sec:introduction}

Instant social video sharing based on the combination of online social networks and user-generated video streaming, has rapidly emerged as one of the most important social media services for users to access contents online \cite{zhang2014understand}. A fundamental reason for the popularity of social video sharing is that it satisfies the users' inherent interests in sharing video contents which are generated and uploaded by users themselves \cite{kwak2010twitter}, with their friends \cite{wasko2005should}. When viewing such user-generated videos, other users need to download the media files from servers over the Internet.

As a result, the placement of instant social video content and the network performance between the servers and the users can significantly affect the service quality of instant social video sharing systems. For this reason, many of them try to use the cloud-based services to deploy their systems, and take full advantage of the elastic and geo-distributed server resource availability in the cloud \cite{hajjat2010cloudward, cheng2011load, wuscaling2012, hu2014community}. 
Since online social network services are generally targeted at a large scale of users distributed at different geographic locations, to satisfy the needs of users, possibly with different network conditions, we may need to allocate servers across many different geographic regions and ISPs, for the sake of achieving better network performance by allocating servers in the proximity of users \cite{liu2012case}.

\begin{figure}[t]
	\centering
		\includegraphics[width=0.8\linewidth]{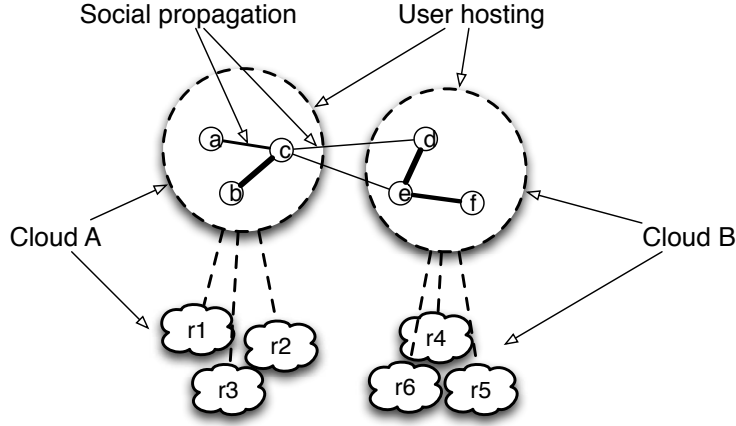}
	\caption{Multi-cloud hosting based on preference- and propagation-aware social graph partition.}
	\label{fig:partition-example}
\end{figure}

\BeginRevision
Today, a number of online multimedia services have been deployed over the geo-distributed cloud and network infrastructure \cite{wuscaling2012}. Intuitively, multi-cloud hosting provides better geographical diversity, since no single cloud provider is able to cover all the regions/ISPs across the Internet \cite{liuoptimizing}, to serve users with their ideal servers. The growing trend of social application as well as the existing geo-distributed deployment for online multimedia applications naturally lead to the idea of {\em multi-cloud instant social video hosting}, or {\em multi-cloud hosting} in short, in that the instant video contents are dispersed to multiple cloud service providers, rather than a single cloud provider. Fig.~\ref{fig:partition-example} gives an example of the multi-cloud hosting (details are to be presented in Sec.~\ref{sec:design}). 
\EndRevision

A fundamental difference between an instant social video sharing system and a traditional content distribution system is the presence of content propagation in the social network, in which social activities such as \emph{sharing} are demanded by users \cite{wasko2005should}. In the context of multi-cloud hosting, besides storage and network cost, social propagation can lead to a large volume of content exchanges between different cloud providers, incurring a high cost of inter-cloud content replication.

The reason is that cloud providers tend to block content replication between the cloud providers with custom tailored pricing schemes: (1) A cloud provider typically encourages a social video sharing system to host user-generated contents, \emph{e.g.}, the \emph{incoming} traffic in Amazon EC2 (Elastic Cloud Computing) is not charged at all; and (2) a cloud provider charges much more for the \emph{outgoing} traffic to a different cloud provider than inside the same cloud, \emph{e.g.}, for the first $10$ TB data transferred from an Virginia EC2 instance outside, the price is $0.12$ USD/GB if the traffic goes to a server hosted by a different cloud provider, while it is only $0.02$ USD/GB if the outgoing traffic remains in Amazon EC2, even though both traffics transfer between the same pair of locations \cite{amazonec2price}.

Such a pricing scheme penalizes outbound transfers, and establishes a roadblock that limits replication across the boundary between different cloud providers, even though such replication is indispensable in the context of social video sharing, since users frequently share and \emph{reshare} content from one another \cite{ye2010measuring}. Taking inter-cloud propagation into account, we seek to study the design space of a multi-cloud hosting strategy that can achieve the following objectives: (1) Satisfying the \emph{cloud-provider preference} of users, so that users are hosted with their ideal cloud providers --- they can share contents to friends and view contents generated by their friends fast \cite{krishnan2009moving}; and (2) Reducing the \emph{cost of inter-cloud traffic} caused by social propagation between users hosted with different cloud providers.

In this paper, we study how to efficiently host an instant social video system with multiple cloud providers based on partition of a propagation-weighted social graph. First, we conduct large-scale measurements to study the benefit of hosting social video contents with multiple cloud providers, the challenges with such multi-cloud hosting, and design guidelines from social propagation characteristics. Second, we formulate the multi-cloud hosting as an optimization problem, which is proven to be NP-hard. \emph{Third}, based on our measurement insights, we propose to solve the problem heuristically by dividing the partition into two phases: an initial preference-aware cloud selection (so that users can upload/download the instant videos to/from their ideal servers), and a \emph{propagation-aware} re-hosting (so as to reduce the cost of replicating the content across the boundary between multiple cloud providers caused by the social propagation). Since only a small set of social connections incurring a large amount of replication cost are re-hosted in our design, the algorithm can efficiently partition large-scale social graphs.

The remainder of this paper is organized as follows. In Sec.~\ref{sec:measure}, we motivate our design by measurement studies of real-world social video sharing and cloud systems. In Sec.~\ref{sec:design}, we formulate the problem and present our multi-cloud hosting design based on the preference- and propagation-aware social graph partition. In Sec.~\ref{sec:evaluation}, we evaluate the performance of our design using trace-driven simulations. In Sec.~\ref{sec:relatedwork}, we discuss related work. Finally, we conclude the paper in Sec.~\ref{sec:conclusion}.

\section{Motivation and Design Principles}
\label{sec:measure}

In this section, we first present our motivation based on measurement results of an instant social video sharing system, then we present our measurement insights for the multi-cloud hosting design for instant social video contents.

\subsection{Assumption: Hosting Users Instead of Contents}

Before presenting the measurement results, we give the assumption made in the multi-cloud hosting. 

Social connections determine how contents propagate between users in the online social network \cite{zhang2014understand}. Content propagation over these social connections turns an online social network to a ``user-subscribing'' network, \emph{i.e.}, each user acts as a source which generates the contents to be subscribed by others \cite{kwak2010twitter}. For this reason, in our study of the multi-cloud hosting of a social video sharing system, we focus on handling the hosting of users, {\em i.e.}, contents generated or shared by a user will be hosted by the same cloud provider assigned to host the user, and different cloud providers are assigned to host different users. 
\BeginRevision
Note that content instead of users is physically stored and served by the cloud servers. The users are actually ``logical'' instances, which generate and propagate contents in the online social networks. These contents can be either static (e.g., photos uploaded by users), and dynamical (e.g., pages generated according to different contexts).
\EndRevision

The benefits of hosting contents of the same user in the same cloud provider are as follows: (1) it avoids individually handling the user-generated contents, which have an extremely large number \cite{facebook-mineral}; and (2) developers for the instant video sharing system can access a user's own contents \emph{locally}, when they are hosted within the same cloud \cite{pujol2010little}.

\BeginRevision
Today, user-level redirection is feasible for several content providers. For example, two different users will use different URLs (associated with their IDs) to download the same content from different servers. As social networks are popularly used by users, such user-aware redirection will be more practical in the future.
\EndRevision

The problem is then to determine which cloud provider is assigned to host which user.

\subsection{Measurement Setup}

To motivate our design, we present the measurement on users' cloud-provider preference, the replication roadblock across the boundary between different cloud providers, and the propagation characteristics in instant social video sharing systems, respectively. We use active and trace-based measurements as follows.

\subsubsection{Instant Video Sharing and Social Propagation} 

We have obtained content upload and request traces from Weishi from the technical team of Tencent, an instant social video sharing system based in China. In Weishi, short videos (in $10$ seconds) are generated by individuals and shared with their ``followers''. Each video in Weishi is transcoded into the following versions: a) 480x480, $2000$ Kbps; b) 480x480, $1050$ Kbps; c) 480x480, $500$ Kbps; d) 480x480, $300$ Kbps. The Weishi traces record two types of user activities in April 2014: (1) Video upload: each record records when a video is generated and uploaded by a particular user; (2) Video download: each item records when a video with a particular version is download by which user, and from which server.

To further study social propagation between users, we have also obtained Weibo traces, containing valuable runtime data of the system in $6$ months (June 2011 --- November 2011). We have collected two types of traces as follows: (1) the social relationship database, which records how users are socially connected to each other at different time points; (2) the microblogs, which are messages posted by the users --- each entry includes the ID, name, IP address of the publisher, time stamp when the microblog is posted, IDs of the parent and root microbloggers if it is a re-post \cite{zhi-tomccap-psar2013}.

\BeginRevision
Weibo and Weishi are different social networks, and different types of online social network services may have different social propagation patterns and characteristics. To jointly use them for studying the service deployment for instant social video service, we carry out the following preprocessing: (1) We choose the traces of Weibo and Weishi, instead of Twitter and Vine, because Weishi and Weibo share a significant fraction of users, as Weishi was developed based on the social graph of Weibo; (2) In Weibo, we only use the propagation traces of videos, and remove other types of multimedia contents.
\EndRevision

\subsubsection{PlanetLab-based Active Measurement}

To practically study the user preference with servers located at different geographic regions over the world, we use PlanetLab-based experiments. We simulate user activities on the PlanetLab nodes, and let them download from and upload to cloud servers allocated from Amazon EC2 (Elastic Cloud Computing) \cite{amazon-ec2}, to study the user preference of different cloud regions.

\subsection{Benefits from Multi-Cloud Hosting}
\label{sec:diverse}

\subsubsection{Diverse Regions/ISPs Improve Service Quality}

To show that diverse server deployment improves the service quality in instance social video sharing, we study the performance of users downloading contents from servers at different geographic locations. In particular, we measure the time users (PlanetLab nodes) spend on downloading contents from the servers allocated at different locations ($7$ Amazon regions are selected). The content size is $1$ MB, and users download the content over HTTP, from the same type of web server. We repeated these download experiments in one week, and calculated the average download speeds of the users, to infer their preference of servers deployed at different regions.

\begin{figure}[t]
	\centering
		\includegraphics[width=0.8\linewidth]{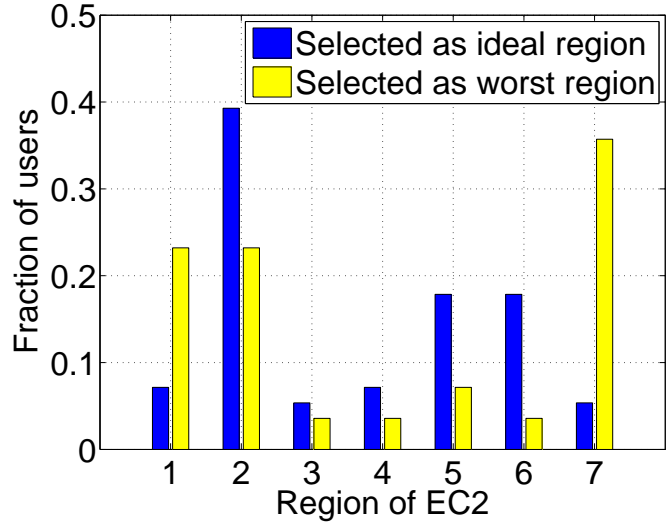}
	\caption{User preference of regions in the Amazon cloud service.}
	\label{fig:best-worst-region}
\end{figure}

Fig.~\ref{fig:best-worst-region} shows the preference of $55$ PlanetLab nodes randomly distributed in different locations over the world. These nodes download the same content from servers deployed at $7$ different regions which are randomly selected from Amazon regions. A pair of bars represent the fraction of each region being selected as the ``ideal'' region (\emph{i.e.}, a PlanetLab node downloads the contents from the region at the highest speed), against the ``worst'' region (\emph{i.e.}, a PlanetLab node downloads the contents from the region at the slowest speed), respectively. We observe that all the regions have an opportunity to be selected as the ideal region by users, indicating that different users have different region preference.

\subsubsection{Multiple Cloud Providers Cover More Service Regions}
\label{sec:amazon-ec2}

Today's cloud providers are scaling their services globally, by building datacenters at different regions and with different ISPs around the world. However, it is difficult for a single cloud provider to cover all the possible regions/ISPs that an instant social video sharing system requires, to serve the users with servers deployed at their ideal regions \cite{liuoptimizing}.

For example, the Amazon EC2 has deployed servers at $9$ regions, including 1.~Virginia, 2.~Oregon, 3.~California, 4.~Ireland, 5.~Frankfurt, 6.~Singapore, 7.~Tokyo, 8.~Sydney, and 9.~Sao Paulo \cite{amazon-ec2}, but these regions fail to locally serve users at some locations, \emph{e.g.}, users in China. On the other hand, some Chinese cloud providers including Tencent Cloud \cite{tencent-cloud} can provide servers in a variety of these regions in China.

It is promising for a social video sharing system to allocate servers from a larger range of regions and ISPs, by utilizing more cloud providers.

\textbf{Measurement insight}. We observe that (1) different cloud providers have deployed servers at different regions/ISPs, and (2) user preference of different regions/ISPs is very different. As a result, multi-cloud hosting of an instant social video sharing system is appealing, since multiple cloud providers allow the system to host contents at a large range of regions/ISPs, so as to improve the possibility for users to download from and upload to their ideal servers.

\subsection{Challenges of Multi-cloud Hosting for Instant Social Video Service}

Next, we study the statistics of content uploads and requests in an instant video sharing service, the replication limitation caused by the pricing schemes of the cloud providers, and the dynamics of social propagation in an instant social video sharing system.

\subsubsection{Instant Video Uploads and Requests}

\begin{figure}[t]
	\begin{minipage}[t]{.48\linewidth}
		\centering
			\includegraphics[width=\linewidth]{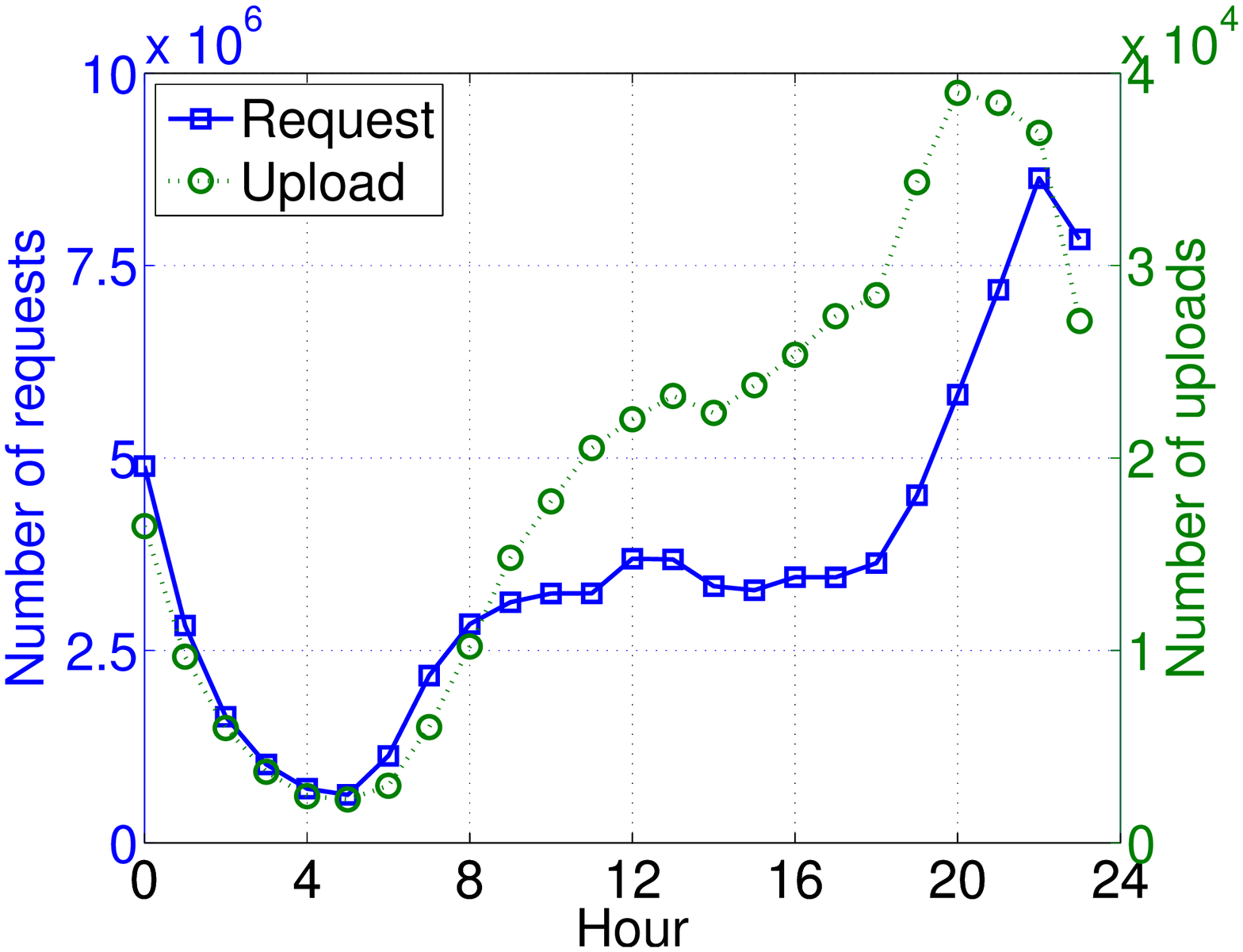}
		\caption{Number of uploads and requests of instance videos over time.}
		\label{fig:upload-request}
	\end{minipage}
	\hfill
	\begin{minipage}[t]{.48\linewidth}
		\centering
			\includegraphics[width=\linewidth]{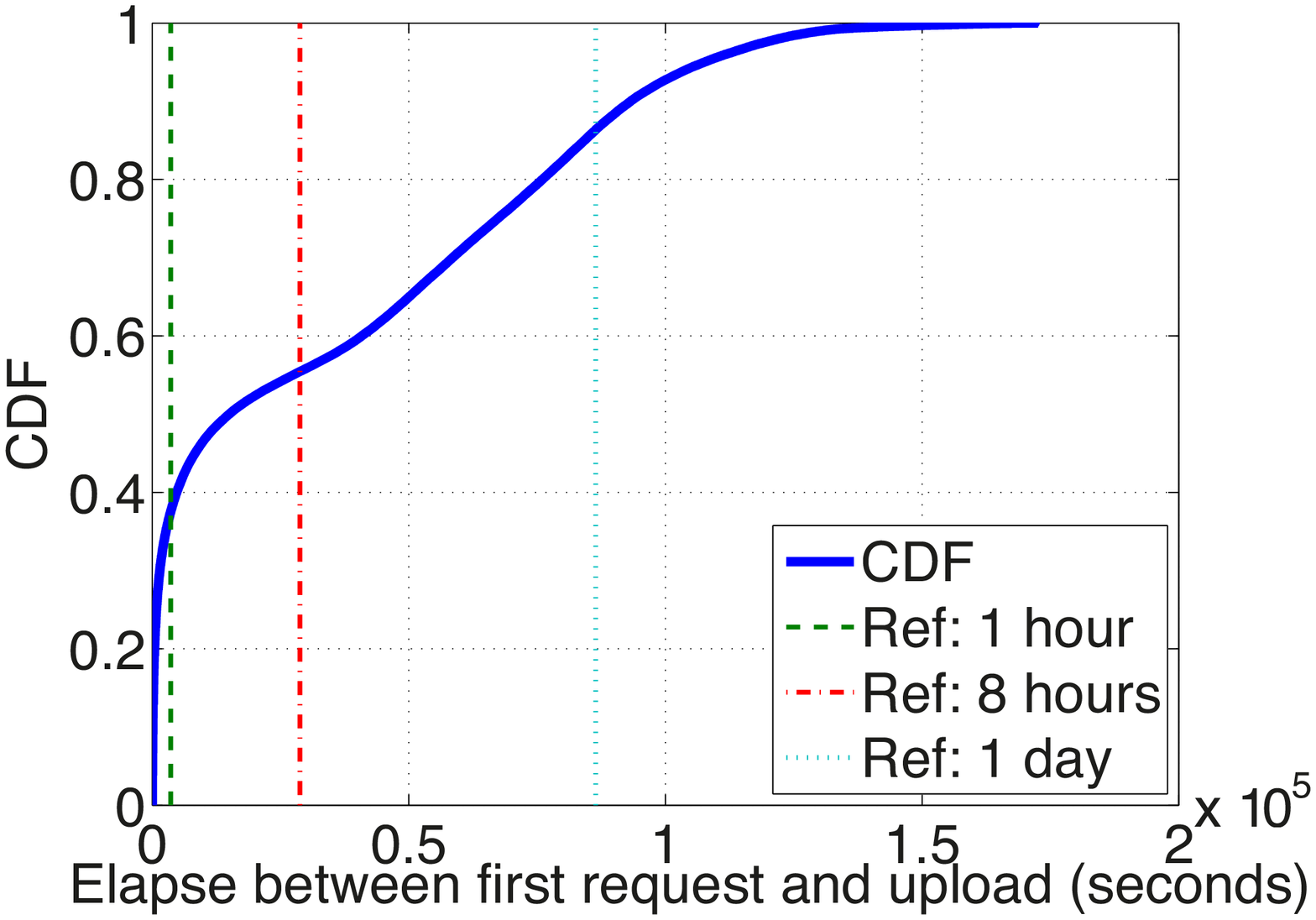}
		\caption{CDF of elapse between upload and the first request.}
		\label{fig:updowntimediff}
	\end{minipage}
\end{figure}

Based on the Weishi traces, we measure the content uploads and requests in an instant social video sharing service. We first study the statistics of the number of video uploads and requests in one day, as illustrated in Fig.~\ref{fig:upload-request}. The two curves in this figure represent the number of instant videos uploaded/requested by users in each time slot ($1$ hour) over time. We observe that both the upload and request curves demonstrate daily patterns, with the peak hours at 8pm and 10pm, respectively. We also observe that the average number of requests is around $100$x larger than that of uploads, indicating that it is likely for the popular videos to be requested by many users, located at different regions. It is thus necessary to deploy these contents into multiple clouds.

We next study the elapse between the upload of a video and the requests. We plot the CDF of the elapse between the upload time of a video and the time the first request for the video was issued in Fig.~\ref{fig:updowntimediff}, over $300,000$ videos. We observe that more than $40\%$ (resp. $55\%$ and $85\%$) of the videos were requested $1$ hour (resp. $8$ hours and $24$ hours) after they were uploaded. These observations show that it is necessary for instant videos to be deployed into multiple clouds timely. 

\subsubsection{Inter-Cloud Replication Cost}

Another challenge is the inter-cloud \emph{replication cost}, which is the cost from replicating contents between cloud providers due to social propagation. In this paper, propagation cost and replication cost are interchangeable. Taking the bandwidth pricing schemes used by Amazon EC2 \cite{amazonec2price} as example, we observe two important pricing schemes used in today's cloud providers as follows. (1) \emph{Incoming content encouragement}. It is observed that cloud providers do not charge the incoming traffic from the Internet, \emph{i.e.}, in the cloud-based social video sharing system, the contents generated by users can be uploaded to servers of any cloud providers for free. (2) \emph{Inter-cloud replication roadblock}. However, it is observed that the outgoing traffic is generally charged. Specifically, the cloud providers charge a regular price of the outgoing traffic when contents are transferred from inside one cloud to another cloud provider; while they charge much less when the outgoing traffic is inside the same cloud. For example, the price scheme for the first $10$ TB outgoing traffic in Amazon EC2 is illustrated in Table \ref{tab:ec2-data-price}. When data is transferred outside an EC2 server at Virginia, the price is $0.12$ USD/GB on average if the traffic goes to a server hosted by a different cloud provider; while it is only $0.02$ USD/GB if the outgoing traffic remains in Amazon EC2.

This pricing scheme restricts the social video sharing systems from freely extending their service to multiple cloud providers, given that contents are indispensably replicated between servers in different cloud providers because of the social propagation between users hosted with different cloud providers.
\BeginRevision
Note that the pricing in our study is actually an input, instead of a mechanism as in typical game theory studies: Our design tries to reduce the replication cost caused by the data transfer price between different cloud providers. 
\EndRevision

\begin{table}[!t]
     \caption{Data transfer price of Amazon EC2 for outgoing traffic (USD/GB for first 1TB, November 2014) \cite{amazonec2price}.}
     \label{tab:ec2-data-price}
     \centering
     \begin{tabular}{|p{0.2\linewidth}||p{0.2\linewidth}|p{0.2\linewidth}|}
          \hline
               \textbf{Region} & \textbf{To another Amazon region} & \textbf{To another cloud provider} \\
          \hline
		  	Virginia	&	0.02 & 	0.12 	\\
			Oregon	& 	0.02 & 	0.12 	\\
			California	& 	0.02 & 	0.12 	\\
			Ireland	& 	0.02 & 	0.12 	\\
			Frankfurt	& 	0.02 & 	0.12 	\\
			Singapore	& 	0.09	&	0.19 \\
			Tokyo	& 	0.09	&	0.201\\
			Sydney	& 	0.14 &	0.19 \\
			Sao Paulo	& 	0.16 &	0.25 \\
		\hline
     \end{tabular}
\end{table}

\subsubsection{Dynamics of Social Propagation}

Another challenge is related to the dynamics of social topology.

\begin{figure}[t]
	\begin{minipage}[t]{.48\linewidth}
		\centering
			\includegraphics[width=\linewidth]{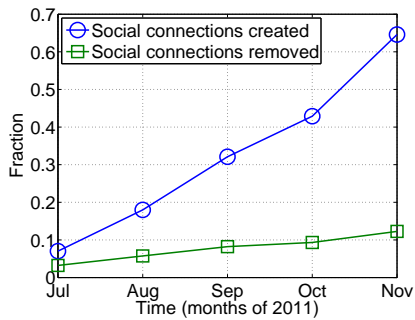}
		\centerline{\parbox[t]{\linewidth}{\scriptsize (a) Social connections created and removed monthly.}}
	\end{minipage}
	\hfill
	\begin{minipage}[t]{.48\linewidth}
		\centering
			\includegraphics[width=\linewidth]{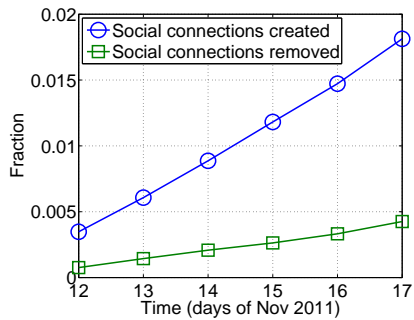}
		\centerline{\parbox[t]{\linewidth}{\scriptsize (b) Social connections created and removed daily.}}
	\end{minipage}     
	\caption{Dynamics of social connections in Tencent Weibo after it was online.}
	\label{fig:conn-dynamics}
\end{figure}

\textit{Creation and removal of social connections}. One year since it was online, social connections within Tencent Weibo were still changing dramatically. Fig.~\ref{fig:conn-dynamics} illustrates the creation and removal ratios of social connections related to a sample of $1$ million users over time (\emph{i.e.}, any social connection that connects at least one user in the $1$ million users is included).

In Fig.~\ref{fig:conn-dynamics}(a), we have a baseline of the number of social connections in June 2011, and each sample in the ``social connection created'' curve represents the creation ratio of social connections since then, while each sample in the ``social connection removed'' curve represents the removal ratio of social connections since then. The creation and removal of social connections among users are relatively dynamical --- in 5 months, over $65\%$ new social connections are created. Besides, much more social connections are created (friending or following others) than removed (un-friending or un-following others). Similar results are also observed in a period of one week in November 2011 in Fig.~\ref{fig:conn-dynamics}(b).

For a newly deployed social video sharing system, the social connections can
change dramatically over time for a long period.

\textbf{Measurement insight}. Multi-cloud hosting of a social video sharing system is challenging, because (1) the cloud providers are using pricing schemes that block the replication of contents between cloud providers, and (2) the social topology is changing dynamically due to frequent creation and removal of the social connections after the social video sharing system is deployed.

\subsection{Principles Learnt from Measurement Studies}
\label{sec:propagation}

We study the characteristics of the social propagation in the online social network, which can guide the multi-cloud hosting design.

\subsubsection{A Few Server Regions Are Enough for Most of the Users}

After a content is generated or shared by a user in the online social network, her friends are the ones who are to view the content. As mentioned above, these friends have different preference of regions to download the contents from. We study how many server regions are required to host a user so that every friend of her can download the content from their ideal regions. Based on the Weibo traces, we retrieve users' geographic locations and estimate their ideal server regions based on the geographic distance, \emph{i.e.}, an ideal server region is one closest to the user.

\begin{figure}[t]
	\begin{minipage}[t]{.48\linewidth}
		\centering
			\includegraphics[width=\linewidth]{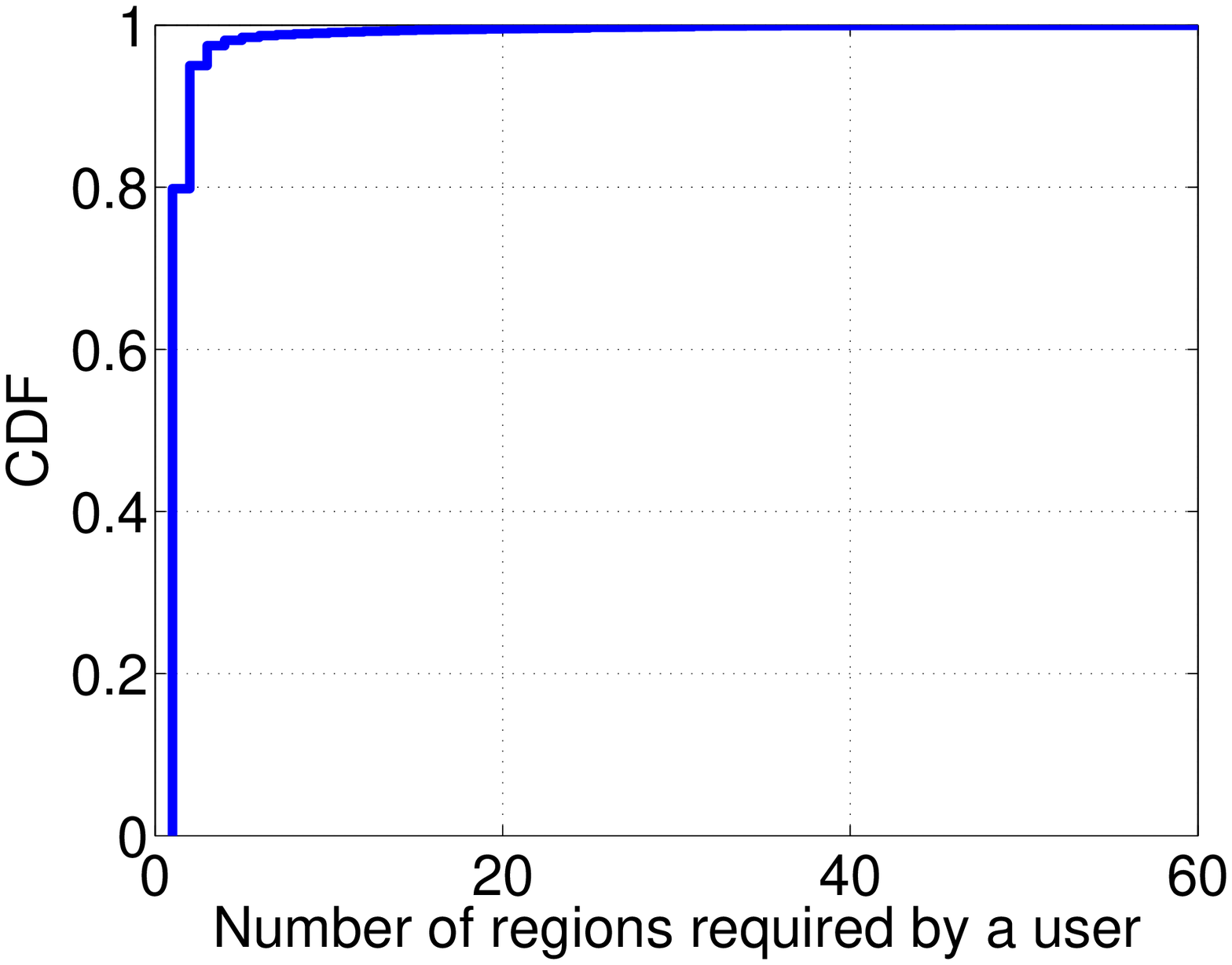}
			\caption{CDF of the number of regions demanded by a user to serve all her friends with their ideal servers.}
			\label{fig:different-region-req}
	\end{minipage}
	\hfill
	\begin{minipage}[t]{.48\linewidth}
		\centering
			\includegraphics[width=\linewidth]{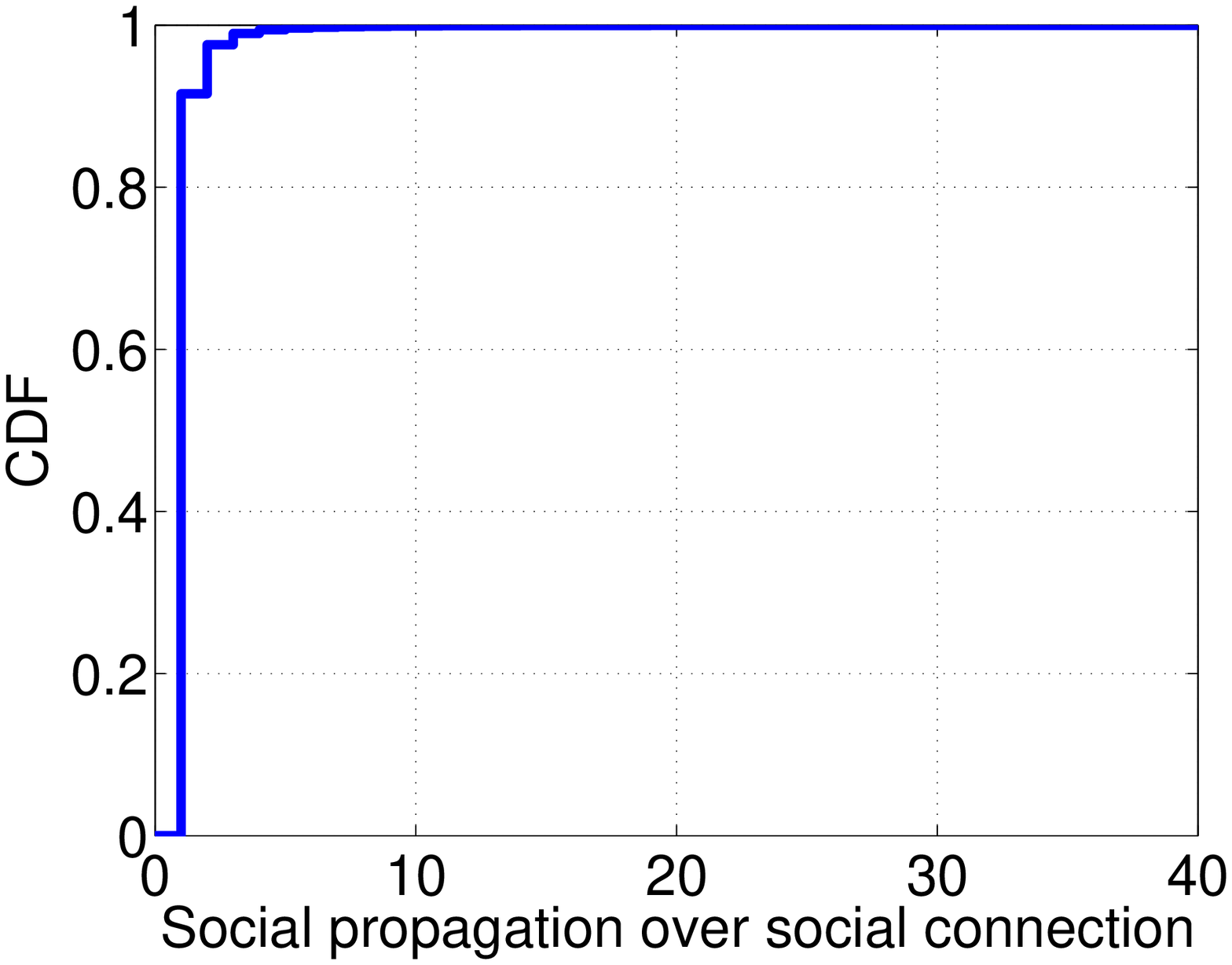}
		\caption{CDF of the propagation weights of social connections.}
		\label{fig:propnum-vs-connrank}
	\end{minipage}     
\end{figure}

\BeginRevision
Fig.~\ref{fig:different-region-req} illustrates the CDF of the number of regions demanded by a user so that all her friends can download the content from their ideal regions. We observe that this number follows a heavy-tailed distribution, \emph{i.e.}, while some of the users need to be hosted with many regions to serve their friends with their ideal regions, most of the users only need a small number of regions, which are highly possible to be covered by a single cloud provider. Since the first type of users tend to be hosted by almost all the cloud providers available to the social video sharing system, in our design, we are focused on the second type of users, to determine which cloud provider is assigned to host which user.
\EndRevision

\subsubsection{A Few Social Connections Incur a Large Amount of Propagation}
\label{sec:measure-propagation}

\BeginRevision
In an online social network, users can reach the content generated by others through the social connections between them. We observe that the number of contents propagating over different social connections can be quite different. In Fig.~\ref{fig:propnum-vs-connrank}, we plot the CDF of the \emph{propagation weight} (\emph{i.e.}, the number of reshares via a particular social connection in $1$ day) of $10,000$ social connections, randomly chosen from all the social connections among all the $1$ million users.
\EndRevision

We observe that the distribution of the propagation weight over different social connections is also heavy-tailed. To reduce the cost of inter-cloud traffic, we need to take social connections with the dominating propagation weight into account when applying the multi-cloud hosting.

\BeginRevision
\textbf{Measurement insight}. We observe that, (1) for most of the users, each individual of them only needs a few number of regions to serve the contents for his followers, which can be provided by a single cloud provider, though multiple cloud providers are needed to cover regions for all of the users; (2) only a few number of social connections incur a large amount of social propagation, which may be the cause of the dominate inter-cloud data transfer cost.
\EndRevision

Based on the measurement insights, we will present our design of the multi-cloud hosting of an instant video sharing media system in Sec.~\ref{sec:design}.

\section{Multi-Cloud Hosting: Detailed Design}
\label{sec:design}

\begin{figure*}[t]
	\centering
		\includegraphics[width=0.85\linewidth]{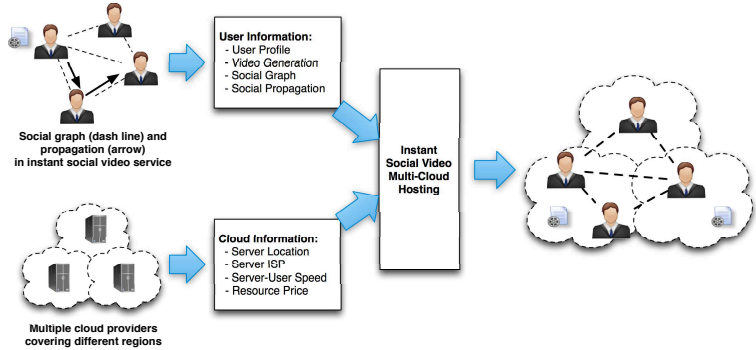}
	\caption{Framework of instant social video multi-cloud hosting.}
	\label{fig:framework}
\end{figure*}

Fig.~\ref{fig:framework} illustrates the framework of our instant social video multi-cloud hosting proposal. We design the instant video content hosting strategies following a data-driven approach. We collect the following information: (1) the social propagation information, including the social graph between users, how they generate videos and how these contents propagate via social connections; and (2) the cloud information, including locations and ISPs of cloud servers, their upload/download speeds to users, and the resource price of these cloud servers (\emph{e.g.}, storage, data). Based on these information, we carry out the multi-cloud hosting strategies, to partition users to different cloud providers, such that users can receive videos with good streaming quality, as well as that the overall content replication cost is minimized.

We need to strategically determine which users should be hosted by which cloud provider, so as to not only satisfy users' cloud-provider preference but also reduce the cost of inter-cloud content replication. In this section, we present our detailed design for instant social video multi-cloud hosting.

Fig.~\ref{fig:partition-example} illustrates the idea of hosting an instant social video sharing system with multiple cloud providers based on the preference- and propagation-aware social graph partition. In this figure, $a,b,\ldots,f$ represent users in the social network, and $r1, r2, \ldots, r6$ represent regions with servers deployed by two cloud providers $A$ and $B$, where $\{r1,r2,r3\}\in A$ and $\{r4,r5,r6\}\in B$. Each user can generate and share a number of contents, which will be downloaded by her friends. The segments between users represent the social connections, which can be retrieved from the online social network. The thickness of a segment represents the propagation level between two users, \emph{i.e.}, a thicker segment indicates that more contents propagate between two users. Recall that such propagation will cause the content replication between the servers where the two users are hosted, as presented in Sec.~\ref{sec:measure}.

We assume that users $a$, $b$, and $c$ prefer cloud $A$, while users
$d$, $e$, and $f$ prefer cloud $B$, \emph{i.e.}, better download and
upload performance can be achieved if they are hosted with their ideal
cloud provider. In this example, we observe that the partition
(indicated by the two large dashed circles) of the users can satisfy the
preference of all users, as well as minimize the inter-cloud
propagation, since the propagation weights of social connections $(c,d)$
and $(c,e)$ are much smaller than that of other social connections.
However, in most of the cases, satisfying users' cloud-provider
preference and minimizing the inter-cloud propagation will conflict
(\emph{e.g.}, when the propagation weight between $c$ and $d$ is very
large), and we need to strategically achieve the two objectives jointly.

Next, we present the formulation of the multi-cloud hosting problem, and
our solution based on the measurement insights.

\subsection{Problem Formulation}

In this subsection, we will formulate the multi-cloud hosting of an instant social video sharing service into an optimization problem. In particular, the objectives we seek to achieve are as follows: (1) we need to satisfy the cloud-provider preference of users who are influenced to download the contents shared by their friends; (2) we need to satisfy the cloud-provider preference of users who generate and upload the contents; and (3) we need to reduce the inter-cloud traffic caused by social propagation between users that are hosted with different cloud providers.

Before we present the details of our design, we summarize some important
notations in Table~\ref{tab:notations}.

\begin{table}[!t]
	\caption{Important notations} 
	\label{tab:notations}
	\renewcommand{\arraystretch}{1.3}
	\centering
	\begin{tabular}{|p{0.12\linewidth}||p{0.78\linewidth}|}
		\hline 
			Symbol & Definition \\ 
		\hline
		\emph{$u,v,w$} 			& Indices for users in the online social network \\
		\emph{$c,d,f$} 			& Indices for cloud providers \\
			\emph{$\graphG=\{\setV,\setE\}$} 			& The social graph with users in set $\setV$ and social connections in set $\setE$\\
			\emph{$\setC$} 				& The set of cloud providers \\
			\emph{$e_{uv}$} 			& The propagation weight of social connection $u \rightarrow v$ \\
			\emph{$\setR_c$} 			& The set of regions in cloud provider $c$ \\
			\emph{$\setF_u$} 			& The set of friends of user $u$ \\
			\emph{$\ch(v,s)$} 			&  The preference level for user $v$ to download contents from servers  at region $s$\\
			\emph{$\ch'(v,s)$} 			&  The preference level for user $v$ to upload contents to servers at region $s$\\
			\emph{$Q(u,c)$} 			& The local download index of user $u$ hosted with cloud provider $c$ \\
			\emph{$W(u,c)$} 			& The local upload index of user $u$ hosted with cloud provider $c$ \\
			\emph{$\pref(u,c)$} 			& The preference of user $u$ to the cloud provider $c$ \\
			\emph{$Y(u,c)$} 			& The cost of inter-cloud replication between user $u$ and her friends if $u$ is hosted with cloud provider $c$\\
			\emph{$p_{cd}$} 			& The data-transfer price of inter-cloud traffic from cloud $c$ to cloud $d$ \\

			\emph{$\alpha$} 			& The parameter used to balance satisfying users' cloud-provider preference and reducing inter-cloud propagation\\
			\emph{$\gain_X$} 			& The gain of re-hosting users using the strategy $X$\\
			\emph{$\rep(\cdot)$} 		& The cost caused by inter-cloud propagation in the re-hosting\\
			\emph{$\eta$} 			& A threshold used to determine which social connections are considered in the re-hosting.\\
		\hline 
	\end{tabular} 
\end{table}

\subsubsection{Social Graph and Multiple Cloud Providers}

Let $\graphG=\{\setV,\setE\}$ denote the social graph with each node $u \in \setV$ representing a user in the online social network, and each edge $e_{uv} \in \setE$ denoting the propagation weight from user $u$ to user $v$. Let $\setC$ denote the set of cloud providers that the instant social video sharing system can be hosted with. Each cloud provider $c \in \setC$ has a set of regions $\setR_c$ where users can be hosted.

\subsubsection{Cloud Preference of a User}

The cloud-provider preference of a user includes the following two
perspectives: (1) improving the download performance by hosting the user
with a cloud so that the user's friends can download from regions that
are close to them; and (2) improving the upload performance by hosting
the user with a cloud with regions that are close to the user himself.

\textit{Local download index}. Let $Q(u,c)$ denote the local download
index of hosting user $u$ with cloud $c$, to satisfy her friends to
download from local regions. $Q(u,c)$ is calculated as follows:
$$
	Q(u,c) = \sum_{v \in \setF_u}  \max_{s \in \setR_c} \ch(v, s),
$$
where $\setF_u$ is the set of user $u$'s friends in the online social
network, $r_v$ is the region where user $v$ is located, and $\ch(v,s)$
denotes the preference level for user $v$ to
download contents that are hosted at region $s$. $\ch(v,s)$ depends on
the network condition between user $v$ and region $s$, and large $\ch(v,s)$
indicates that better network performance can be achieved in the
download. The rationale of $Q(u,c)$ is that hosting a user $u$ with a
cloud $c$ with large $Q(u,c)$ can benefit her friends, who can download
the contents generated or shared by user $u$ from their ideal cloud
regions.

\textit{Local upload index}. We also seek to find an ideal server for
the user himself to upload the generated contents to. Let $W(u,c)$
denote the local upload index, which represents the upload performance
achieved at user $u$ when $u$ is hosted with cloud $c$. $W(u,c)$ is
defined as follows:
$$
	W(u,c) = \max_{s \in \setR_c} \ch'(u, s) K_u,
$$
where $\ch'(u,s)$ denotes the preference level for user $u$ to upload
his generated contents to a server at region $s$, and $K_u$ is the
average amount of content that can be generated by user $u$ in the next
time slot. A larger local upload index $W(u,c)$ indicates that better
upload gain can be achieved when $u$ is hosted with the cloud provider
$c$. 

In our experiments, $\ch(v,s)$ and $\ch'(v,s)$ can be estimated either by the geographic distance between the user and the server region, or using the historical network performance.

\textit{Cloud-provider preference of a user}.  In our design, the overall
cloud-provider preference of a user takes both the local download
performance (for the user's friends) and local upload performance (for
the user himself) into consideration. The overall cloud-provider
preference of a user is then the combination of the two indices. We
denote $\pref(u,c)$ as user $u$'s preference of cloud provider $c$,
defined as follows:
\begin{equation}
	\pref(u,c) = Q(u,c) + \beta_u W(u,c),
	\label{eq:pref}
\end{equation}
where $\beta_u$ is an implementation parameter used to combine the two indices, depending on the characteristics of a user, \emph{e.g.}, a large $\beta_u$ for a user who frequently generates and uploads contents from a mobile device, so that a cloud provider with servers the user can upload content fast to will have a larger preference index $\pref(u,c)$ with the user.

\subsubsection{Replication Cost Due to Inter-Cloud Propagation}

As a unique cost in the multi-cloud hosting, the inter-cloud traffic
cost is caused by the social propagation between users that are hosted
with different cloud providers, due to the pricing scheme of the cloud
providers we have shown in Sec.~\ref{sec:measure}. In the online social
network, the common social activities such as sharing contents
\cite{wasko2005should} make the contents associate with different users
dynamically.

Due to the high cost of replicating contents from servers in one cloud
to servers in another cloud, we need to take the cost of inter-cloud
content replication caused by social propagation into account. We define
the replication cost for user $u$ hosted with cloud $c$ as follows: 
$$
	Y(u,c) = \sum_{v \in \setF_u} {\frac{1}{2} (p_{C_uC_v}e_{uv} + p_{C_vC_u}e_{vu})},
$$
where $p_{cd}$ is the data transfer price for replicating a content from cloud $c$ to cloud $d$, \emph{i.e.}, the price that the instant social video sharing system has to pay when contents are replicated between two cloud providers rather than inside one cloud provider. $C_u$ is the cloud with which user $u$ is hosted. $p_{C_uC_v}e_{uv}$ is the cost of inter-cloud traffic of social connection $(u,v)$, depending on the actual pricing scheme used by cloud providers. Next, we will discuss how these objectives are achieved in our multi-cloud hosting design.

\subsubsection{Problem Formulation and Analysis}

\textit{Optimization}. 
To satisfy users' cloud-provider preference as well as reducing the
inter-cloud propagation, the multi-cloud hosting can be formulated as
an optimization problem by combining the two objectives, as follows:
\begin{equation}
	\max_{\{C_u|u \in \setV\}} \sum_{u \in \setV} \alpha \pref(u,C_u) - (1-\alpha) \gamma Y(u,C_u),
	\label{eq:obj}
\end{equation}
subject to:
$$
	C_u \in \setC, \forall u \in \setV,
$$
where $C_u$ is the objective variable that determines the cloud provider
with which user $u$ is hosted, $\gamma$ is the parameter used to combine
the inter-cloud \BeginRevision  \emph{propagation cost} (The cost caused by content replication across different cloud servers due to social propagation) \EndRevision with the users' cloud-provider
preference, and $\alpha$ is the parameter to adjust the weight between the
two objectives. The optimization variables give the choices of cloud
providers for users in the online social network.

\textit{Proof of NP-Hardness}. The optimization to determine the
multi-cloud hosting is NP-hard in general. 

\begin{proof} To prove this, we reduce a MCP (Multiterminal Cut Problem)
\cite{dahlhaus1994complexity}, which is NP-hard, to it. In the MCP, we
are given an edge-weighted graph $G$ and a subset of $k$ vertices called
terminals, and asked for a minimum weight set of edges that separates
each terminal from all the others. Next, we show that the MCP can be
reduced to the multi-cloud hosting problem. We build a social graph $G'$
which has the same structure with $G$. The reduction is as follows. (1)
We have $k$ cloud providers $c_1,c_2,\ldots, c_k$ for the multi-cloud
hosting, and the data-transfer price between any two cloud providers is
$1$. (2) We let the propagation weight of a social connection in $G'$ be
the same as the edge weight of the corresponding edge in $G$. (3)
Without loss of generality, we let users $u_1,u_2,\ldots,u_k$ in $G'$ be
the ones corresponding to the $k$ terminals in $G$, and we assign the
cloud-provider preference of users as follows:
$$
	\begin{cases}
		\pref(u_i,c_j) = 0, & i > k, j = 1,2,\ldots,k \\
		\pref(u_i,c_j) = 0, & j \ne i, i = 1,2,\dots,k\\
		\pref(u_i,c_j) = L, & j = i, i = 1,2,\ldots,k \\
	\end{cases},
$$
where $L$ is a const cloud-provider preference, which can be assigned
with a value large enough (\emph{e.g.}, the sum of all propagation
weight) so that every user $u_i, i = 1,2,\ldots,k$ has to be hosted with
the cloud $c_i$ to achieve the optimal multi-cloud hosting. Thus, the
$k$ users will be separated from each other in the multi-cloud hosting
(they are hosted with different cloud providers respectively) while the
overall social propagation is minimized. If the multi-cloud hosting
problem can be solved, then the solution of the original MCP can be
achieved as well, \emph{i.e.}, the set of edges corresponding to the
social connections between any two cloud providers. Thus, the
multi-cloud hosting problem is NP-hard.  \end{proof}

\subsection{Heuristic Multi-Cloud Hosting}

Based on our measurement insights, we design a heuristic algorithm to
solve the multi-cloud hosting problem.
Algorithm~\ref{alg:multi-cloud-hosting} presents our strategy to
partition the social graph, determining which users should be hosted
with which cloud providers. Our algorithm includes two phases as
follows: (1) In the initial preference-aware cloud selection (lines
\ref{ln:init-cloud-sel-start} -- \ref{ln:init-cloud-sel-end}), a user is
assigned to a cloud provider according to only the cloud-provider
preference of hosting a user ($\pref(u,c)$), without considering the
social propagation; (2) In the propagation-aware re-hosting (lines
\ref{ln:re-host-user-start} -- \ref{ln:re-host-user-end}), pairs of
users are assigned with different cloud providers to reduce the
inter-cloud propagation. We will present the two phases as follows.

\begin{algorithm}[t]
\caption{Heuristic multi-cloud hosting algorithm}\label{alg:multi-cloud-hosting}
\begin{algorithmic}[1]
	\Procedure{Multi-Cloud Hosting}{} 
		\ForAll{$u \in \setV$}  \label{ln:init-cloud-sel-start}
			\State $C_u \gets \arg\max_{c \in \setC} {\pref(u,c)}$  \label{ln:sel-max}
		\EndFor \label{ln:init-cloud-sel-end}
		\State Sort social connections in their propagation weight's descending order \label{ln:re-host-user-start}
		\Repeat
			\State $\{u,v\}, C_u \ne C_v \gets$ head of the sorted list
			\State Calculate $\gain_A$, $\gain_B$, $\gain_C$ and $\gain_D$ in Eq.~(\ref{equ:pref})
			\State $\gain'_X = \max\{\gain_A, \gain_B, \gain_C, \gain_D\}$ 
			\If{$X \in \{B,C,D\}$}
				\State Re-host user $u$ and user $v$ accordingly (Sec.~\ref{sec:design-rehosting})
			\EndIf
			\State Remove social connection $\{u,v\}$ from the sorted list
		\Until{$\gain'_X < \eta$} \label{ln:re-host-user-end}
	\EndProcedure 
\end{algorithmic}
\end{algorithm}

\subsubsection{Preference-Aware Initial Hosting}

In this phase, an initial cloud provider is assigned to host each user, only according the cloud-provider preference ($\pref(u,c)$) of users. For each user in the online social network, the ideal cloud provider is the one that can maximize its cloud-provider preference among all available cloud providers (line~\ref{ln:sel-max}). After the initial cloud selection, users are assigned to the cloud providers that can maximize their preference; however, such assignment can result in a large cost of inter-cloud propagation ($Y(u,c)$). Next, users are adjusted to reduce the inter-cloud propagation.

\subsubsection{Propagation-Aware Re-Hosting}
\label{sec:design-rehosting}

Our re-hosting strategy is to change the cloud providers for users so that the inter-cloud replication cost can be reduced. The re-hosting procedure works as follows.

First, the social connections are ranked between users that are not hosted with the same cloud in the descending order of the propagation weight (line \ref{ln:re-host-user-start}). A pair of users who have a large propagation weight between them are can be hosted with the same cloud to reduce the inter-cloud replication cost.

\begin{figure}[t]
	\begin{minipage}[t]{.48\linewidth}
		\centering
			\includegraphics[height=\figureheightfour]{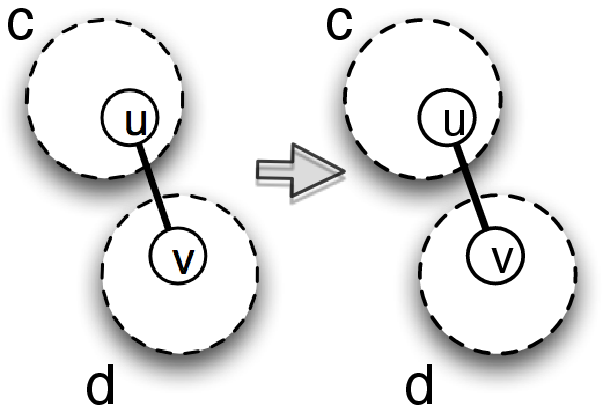}
		\centerline{{\scriptsize (A)}}
	\end{minipage}
	\hfill
	\begin{minipage}[t]{.48\linewidth}
		\centering
			\includegraphics[height=\figureheightfour]{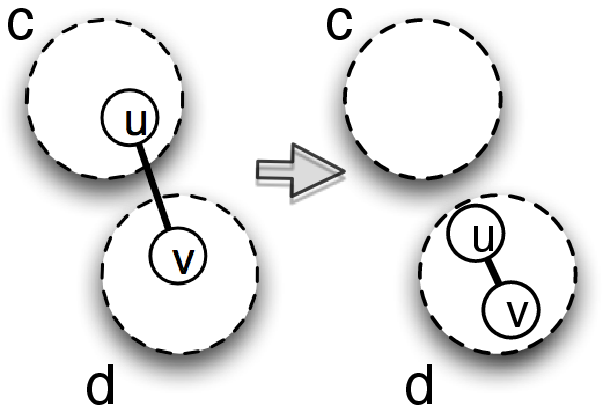}
		\centerline{{\scriptsize (B)}}
	\end{minipage}
	\begin{minipage}[t]{.48\linewidth}
		\centering
			\includegraphics[height=\figureheightfour]{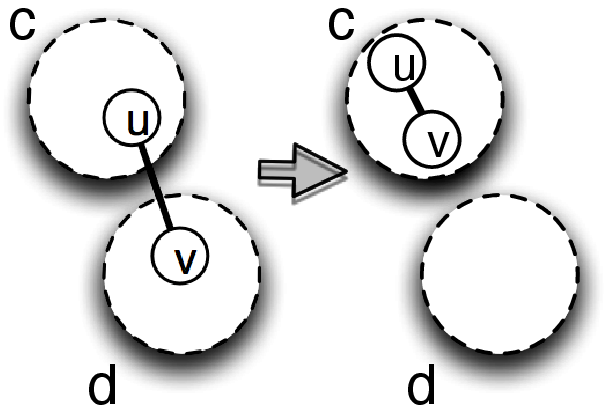}
		\centerline{{\scriptsize (C)}}
	\end{minipage}
	\hfill
	\begin{minipage}[t]{.48\linewidth}
		\centering
			\includegraphics[height=\figureheightfour]{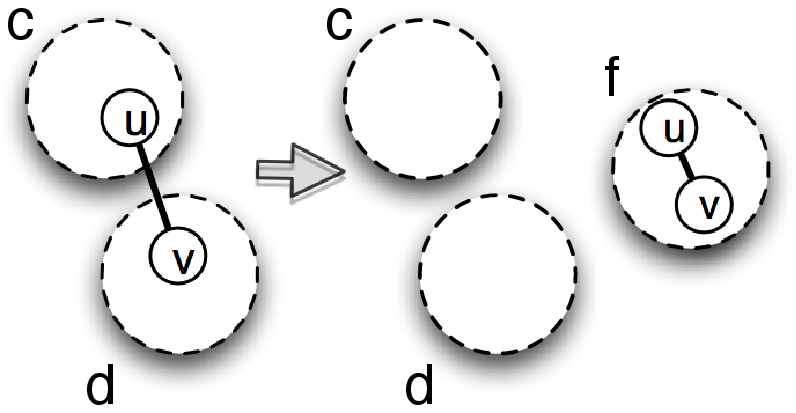}
		\centerline{{\scriptsize (D)}}
	\end{minipage}
	\caption{Strategies of re-hosting a pair of users.}
	\label{fig:re-hosting}
\end{figure}

Second, we present the re-hosting approach. Let $u$ and $v$ denote the pair of users whose social connection has the largest propagation weight. Fig.~\ref{fig:re-hosting} illustrates the schemes that we can apply to improve the partition: (1) keep the original hosting strategy, as illustrated in Fig.~\ref{fig:re-hosting}(A); (2) re-host $u$ to the cloud by which $v$ is hosted, as illustrated in Fig.~\ref{fig:re-hosting}(B); (3) re-host $v$ to the cloud by which $u$ is hosted, as illustrated in Fig.~\ref{fig:re-hosting}(C); and (4) re-host both $u$ and $v$ to a new cloud provider $f$, as illustrated in Fig.~\ref{fig:re-hosting}(D). We can improve the performance by applying one among the four strategies, to re-host $u$ and $v$ with one cloud.

\emph{Third}, to determine which strategy is used to re-host user $u$
and user $v$, we use the following heuristic: we define a gain $\gain_X,
X \in \{A,B,C,D\}$ for each re-hosting scheme. $\gain_A$ is the gain of
hosting user $u$ with cloud $c$, and user $v$ in cloud $d$, without any
change to the original hosting (it is the baseline); $\gain_B$ is the
gain of hosting users $u$ and $v$ with cloud $d$; $\gain_C$ is the gain
of hosting users $u$ and $v$ with cloud $c$; and $\gain_D$ is the gain
of hosting users $u$ and $v$ with a new cloud provider $f$. The gain
is defined as follows:
\begin{equation}
	\begin{array}{l}
		\begin{array}{l}
			\gain_A = 0 \\
		\end{array}\\
		\begin{array}{l}
			\gain_B = \alpha[\pref(u,d) - \pref(u,c)] - (1-\alpha)\rep(u \rightarrow d) \\
		\end{array}\\
		\begin{array}{l}
			\gain_C = \alpha[\pref(v,c) - \pref(v,d)] - (1-\alpha)\rep(v \rightarrow c) \\
		\end{array}\\
		\begin{array}{l}
			\gain_D = \max_{f \ne c,d} \{\alpha[\pref(u,f) + \pref(v,f) - \pref(u,c) \\
			 - \pref(v,d)] - (1-\alpha)\rep(u,v \rightarrow f) \}
		\end{array}
	\end{array},
	\label{equ:pref}
\end{equation}
where the first part ($\alpha[\cdot]$) is the gain of improving the
cloud-provider preference of user $u$ and $v$ according to the
re-hosting, and the second part ($-(1-\alpha)\rep(\cdot)$) represents
the gain by reducing the inter-cloud replication cost under different
re-hosts. $\rep(\cdot)$ is defined as follows:
\begin{eqnarray}
\begin{cases}
		\begin{array}{l}
			\frac{1}{2}\sum_{w|C(w) = c} {\gamma (p_{cd} e_{wu} + p_{dc} e_{uw})} \\
		  	- \frac{1}{2}\sum_{w|C(w) = d} {\gamma (p_{cd} e_{uw} + p_{dc} e_{wu})} 
		\end{array} &  u \rightarrow d\\
		\begin{array}{l}
			\frac{1}{2}\sum_{w|C(w) = d} {\gamma (p_{dc} e_{wv} + p_{cd} e_{vw})} \\
		  	- \frac{1}{2}\sum_{w|C(w) = c} {\gamma (p_{dc} e_{vw} + p_{cd} e_{wv})} 
		\end{array} &  v \rightarrow c \\		
		\begin{array}{l}
			\frac{1}{2}\sum_{w|C(w) = c} {\gamma (p_{cd} e_{wu} + p_{dc} e_{uw})} \\
		  	- \frac{1}{2}\sum_{w|C(w) = f} {\gamma (p_{cf} e_{uw} + p_{fc} e_{wu})} \\
				+ \frac{1}{2}\sum_{w|C(w) = d} {\gamma (p_{dc} e_{wv} + p_{cd} e_{vw})} \\
			  	- \frac{1}{2}\sum_{w|C(w) = f} {\gamma (p_{df} e_{vw} + p_{fd} e_{wv})} \\
		\end{array} &  u,v \rightarrow f \\
	\end{cases}.
	\nonumber
\end{eqnarray}
The re-hosting is performed according to the value of the gains.
Among $\gain_A$, $\gain_B$, $\gain_C$ and $\gain_D$, if $\gain_A$ is the
largest, $u$ will remain in cloud $c$ and $v$ in cloud $d$; if
$\gain_B$ is the largest, $u$ will be hosted with cloud $d$; if $\gain_C$
is the largest, $v$ will be hosted with cloud $c$; otherwise, $v$ will be hosted 
and $u$ with a new cloud provider $f$, which can maximize the re-hosting
gain.

\emph{Fourth}, in our heuristic re-hosting, social connections with the largest propagation weight are first processed. In each pass, we only consider the social connections that can incur a high cost of inter-cloud traffic (as observed in our measurement study, the fraction of such social connections is very small). A threshold $\eta$ is used to determine which social connections are considered in the re-hosting. In our experiments, $\eta$ is selected to terminate the re-hosting loop when the fraction of social connections touched exceeds $20\%$ of all the social connections. The rationale is that, according to our measurement insight in Sec.~\ref{sec:measure-propagation}, in a real online social video sharing system, only the most ``active'' social connections will affect the replication cost.

\BeginRevision
According to our design, the complexity of the algorithm is $\mathcal{O}(\setE|\log|\setE|)$, since only the most influential social connections are considered in our algorithm. In a social network, as $|\setV|$ is similar to $|\setE|$ \cite{caldarelli2007scale}, the complexity is then similar to $\mathcal{O}(|\setV|\log|\setV|)$. 
\EndRevision
\section{Experimental Results}
\label{sec:evaluation}

In this section, we evaluate the performance of our design using simulation experiments driven by the Weibo traces. In our experiments, we will study the satisfaction of users' cloud-provider preference, the reduction of replication cost caused by inter-cloud social propagation, and the efficiency of the heuristic algorithm.

\subsection{Experiment Setup}

\textit{Users and social graph}. We have used a sample of $20,000$ users selected from the social graph of Tencent Weibo, by a BFS-based collection from $10$ random seed users, \emph{i.e.}, we initialize a user set with the $10$ seed users, and iteratively add friends of users that are already in the set, until the size of the set reaches $20,000$ or it is self-contained.

\textit{Content generation and propagation}. We also use the traces of Tencent Weibo to drive the experiments. The actions of posting microblogs in the traces are considered as generating new contents, and the actions of sharing microblogs are used to weight the social propagation between the users. Contents generated and shared need to be hosted with the cloud providers. Based on the traces, we perform the social graph partition for the multi-cloud hosting.

\textit{Cloud regions and prices}. In these Weibo traces, about $80$ regions are observed to have users located in. We have randomly assigned each of these regions to one of $6$ cloud providers. We assume the cloud providers have the same pricing scheme: (1) the price of outgoing traffic to a different cloud provider is $1$, and the price of outgoing traffic to the same cloud provider is $0$, \emph{i.e.}, the roadblock of inter-cloud replication; and (2) the price of incoming traffic is $0$, \emph{i.e.}, the encouragement of incoming contents from the Internet.

According to our design in Sec.~\ref{sec:design}, the locations of users are used to calculate their preference of cloud providers, according to Eq.~\ref{eq:pref}. 

Next, we present the results in our evaluation.

\subsection{Performance Evaluation}

\subsubsection{Satisfying Users' Cloud Preference}

First, we study how users' cloud-provider preference is satisfied in our design. In our experiments, a user's preference of a cloud provider is normalized, \emph{i.e.}, the sum of the preference of all the cloud providers is $1$. We denote the normalized preference as $\npref(u, C_u)$, and use the overall satisfaction of user preference ($\sum_u \npref(u, C_u)$) as the metric to evaluate the performance.

\begin{figure}[t]
	\begin{minipage}[t]{.48\linewidth}
		\centering
			\includegraphics[width=\linewidth]{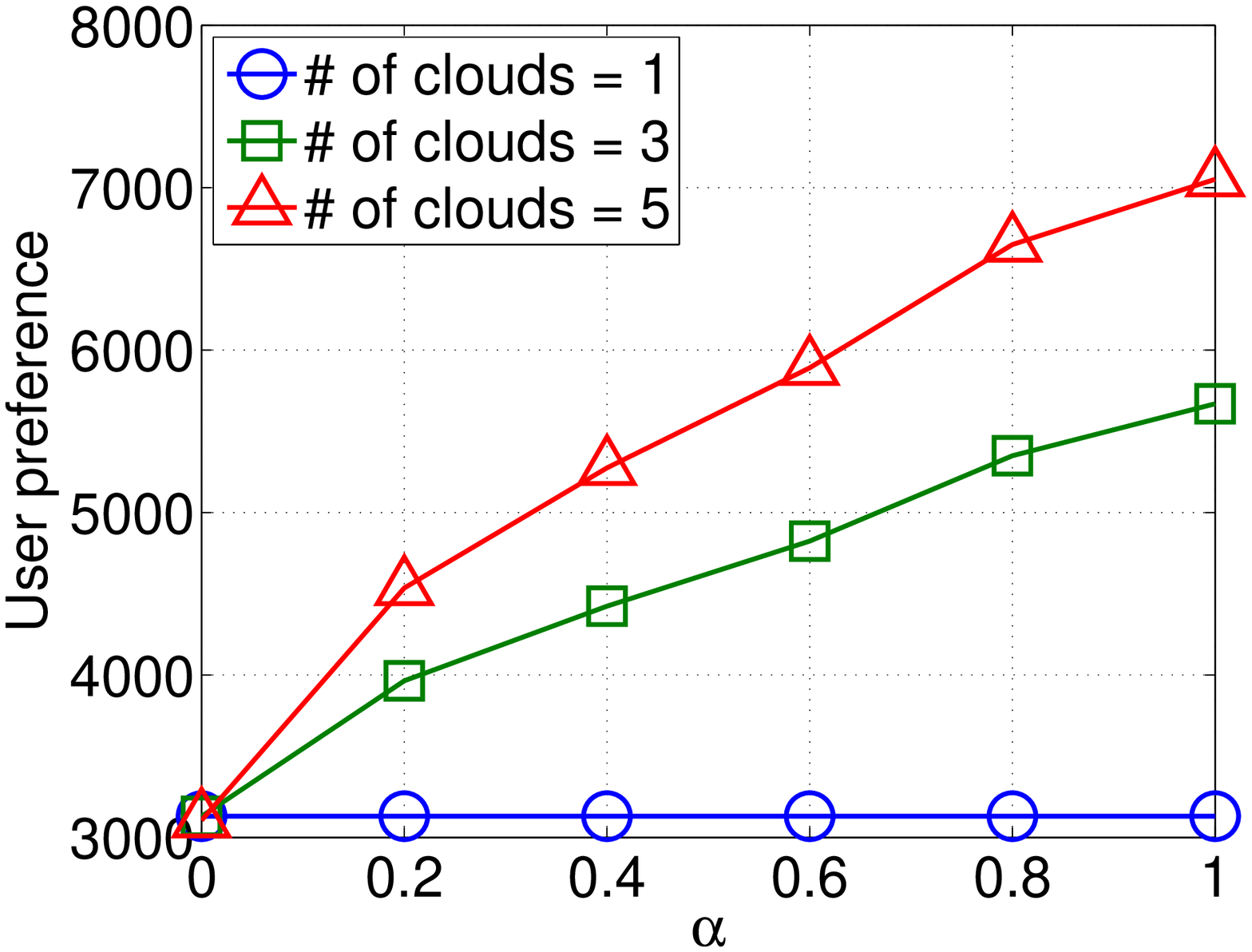}
		\caption{Satisfaction of users' cloud-provider preference.}
		\label{fig:user-pref}
	\end{minipage}
	\hfill
	\begin{minipage}[t]{.48\linewidth}
		\centering
			\includegraphics[width=\linewidth]{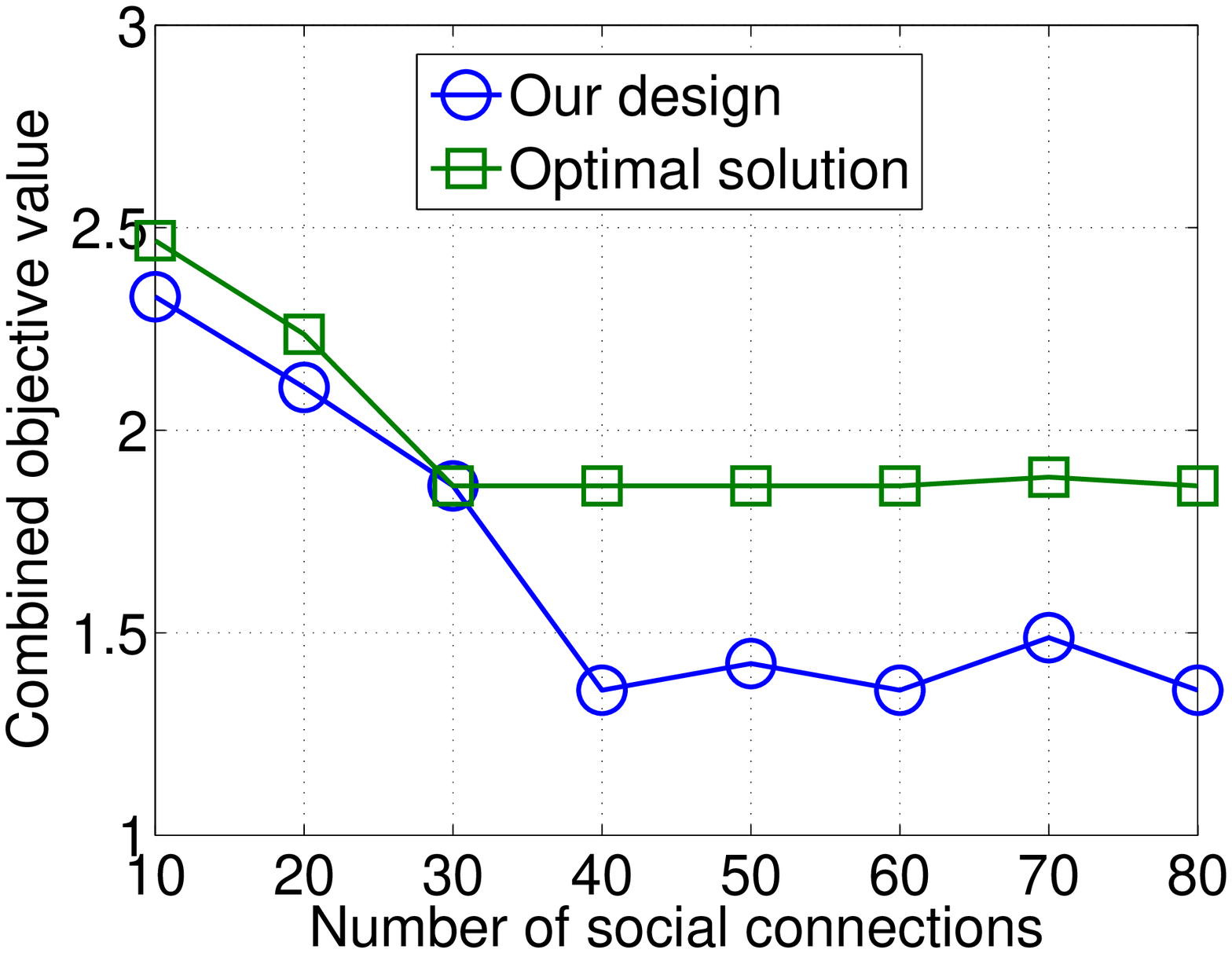}
		\caption{Comparison between our algorithm and the brute-force algorithm with the optimal solution.}
		\label{fig:optimal-cmp}	
	\end{minipage}
\end{figure}

We study the satisfaction of user preference under different weight of parameter $\alpha$. Fig.~\ref{fig:user-pref} illustrates the overall cloud-provider preference versus $\alpha$. Different curves are generated under different numbers of available cloud providers (out of all the $6$ cloud providers) for the multi-cloud hosting. We observe a general increase of users' cloud-provider preference as $\alpha$ grows using more than $1$ cloud providers, since the users' cloud-provider preference is more considered when $\alpha$ is larger. We also observe that when more cloud providers are available for the social video sharing system to choose from, the overall cloud-provider preference can be improved.

\subsubsection{Reducing Inter-Cloud Propagation}

Next, we study the cost caused by inter-cloud propagation. To evaluate the cost, we define two metrics: (1) the propagation cost of all inter-cloud social connections calculated as $\sum_{u,v} p_{C_uC_v} e_{uv}$; and (2) the number of social connections that connect users hosted with different cloud providers.

\begin{figure}[t]
	\begin{minipage}[t]{.48\linewidth}
		\centering
			\includegraphics[width=\linewidth]{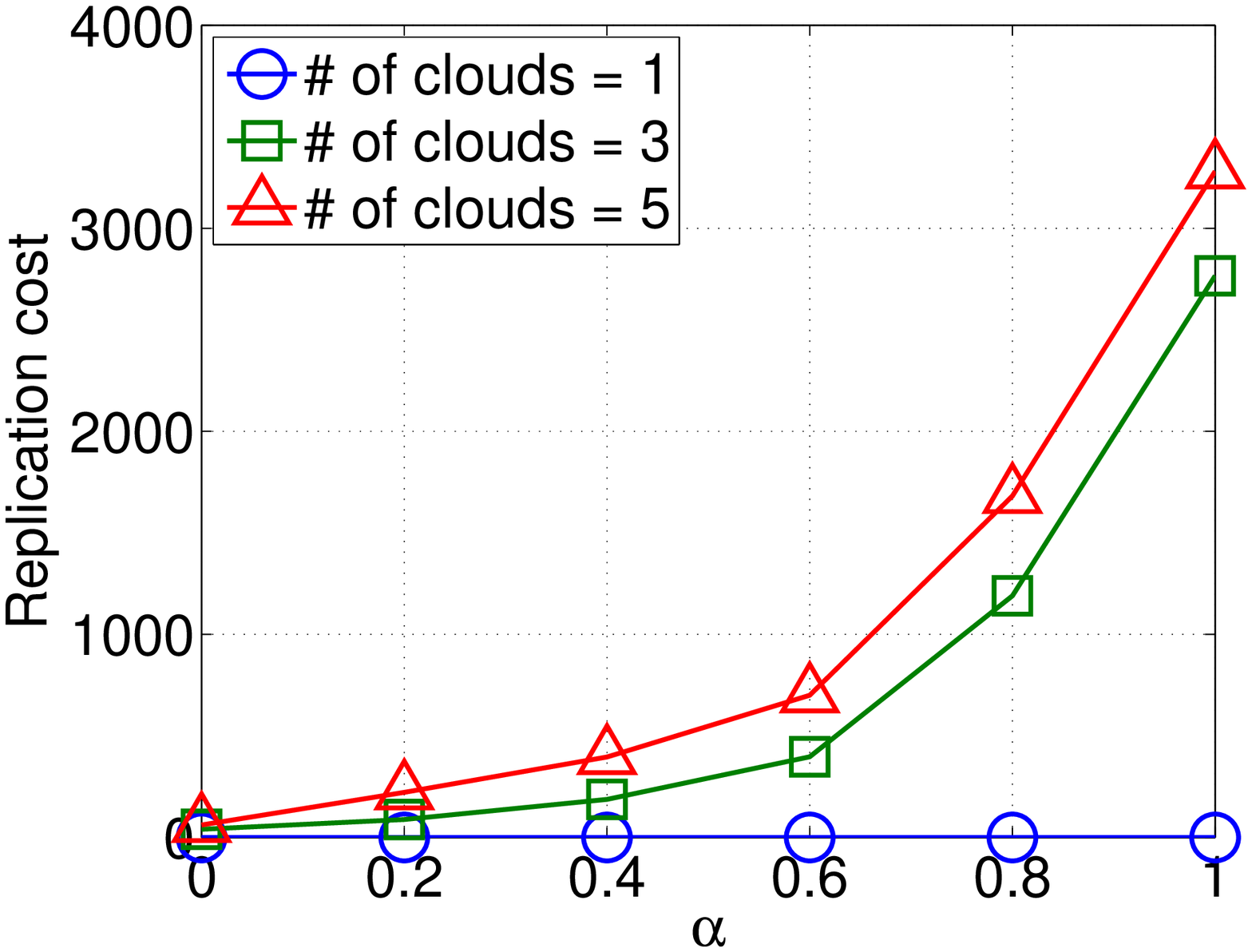}
		\centerline{\parbox[t]{\linewidth}{\scriptsize (a) Replication cost
		versus $\alpha$.}}
	\end{minipage}
	\hfill
	\begin{minipage}[t]{.48\linewidth}
		\centering
			\includegraphics[width=\linewidth]{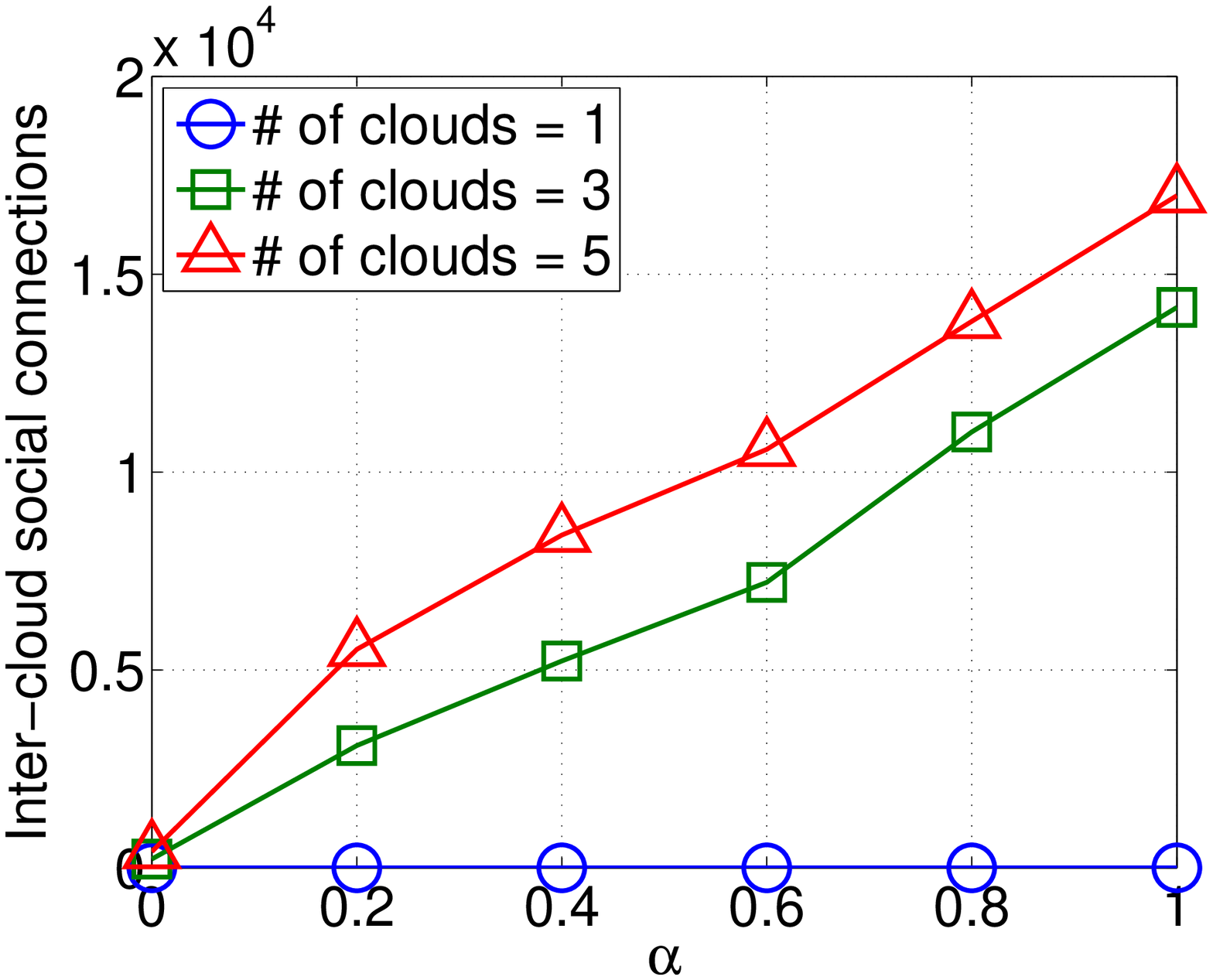}
		\centerline{\parbox[t]{\linewidth}{\scriptsize (b) Number of inter-cloud social connections versus $\alpha$.}}
	\end{minipage}     
	\caption{Replication cost caused by inter-cloud social propagation.}
	\label{fig:inter-cloud}
\end{figure}

In Fig.~\ref{fig:inter-cloud}, we study the impact of $\alpha$ on the cost of inter-cloud propagation. The curves in Fig.~\ref{fig:inter-cloud}(a) illustrate the propagation cost against $\alpha$. We observe that the inter-cloud replication cost generally increases as $\alpha$ grows, since the weight of the inter-cloud propagation becomes smaller in the social graph partition. We also observe that the cost increases much faster when $\alpha$ is larger. This result indicates that strategies only considering users' cloud-provider preference can incur a high inter-cloud propagation cost for the social video sharing system.

The curves in Fig.~\ref{fig:inter-cloud}(b) illustrate the numbers of the inter-cloud social connections. We observe that more social connections span different cloud providers when $\alpha$ grows, and the increase speed of the number of inter-cloud social connections as $\alpha$ increases is much more linear than the propagation cost. The reason is that our algorithm tends to divide friends with small propagation weight between them into different cloud providers, but keep them in the same cloud if the propagation weight is large.

\subsubsection{Performance Comparison}

In our experiments, we have also compared our algorithm (with $\alpha=0.5$) with the following strategies: (1) Random partition, in which users are randomly assigned to the cloud providers; (2) Min-propagation, in which users are partitioned to minimize the inter-cloud propagation; and (3) Max-preference, in which users are hosted with their ideal cloud providers.

\begin{figure}[t]
	\begin{minipage}[t]{.48\linewidth}
		\centering
			\includegraphics[width=\linewidth]{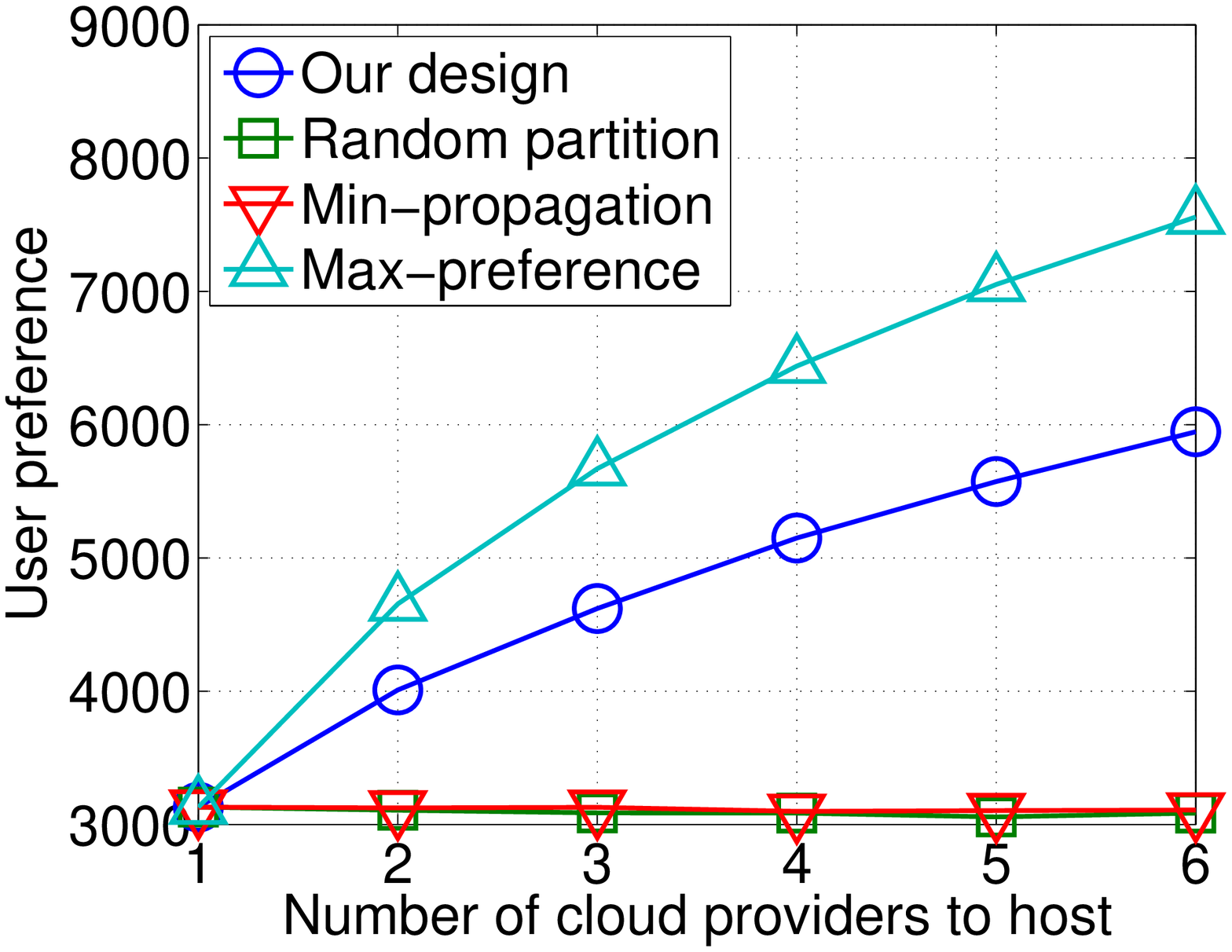}
		\centerline{\parbox[t]{\linewidth}{\scriptsize (a) User preference versus the number of cloud providers to host the instant social video sharing system.}}
	\end{minipage}
	\hfill
	\begin{minipage}[t]{.48\linewidth}
		\centering
			\includegraphics[width=\linewidth]{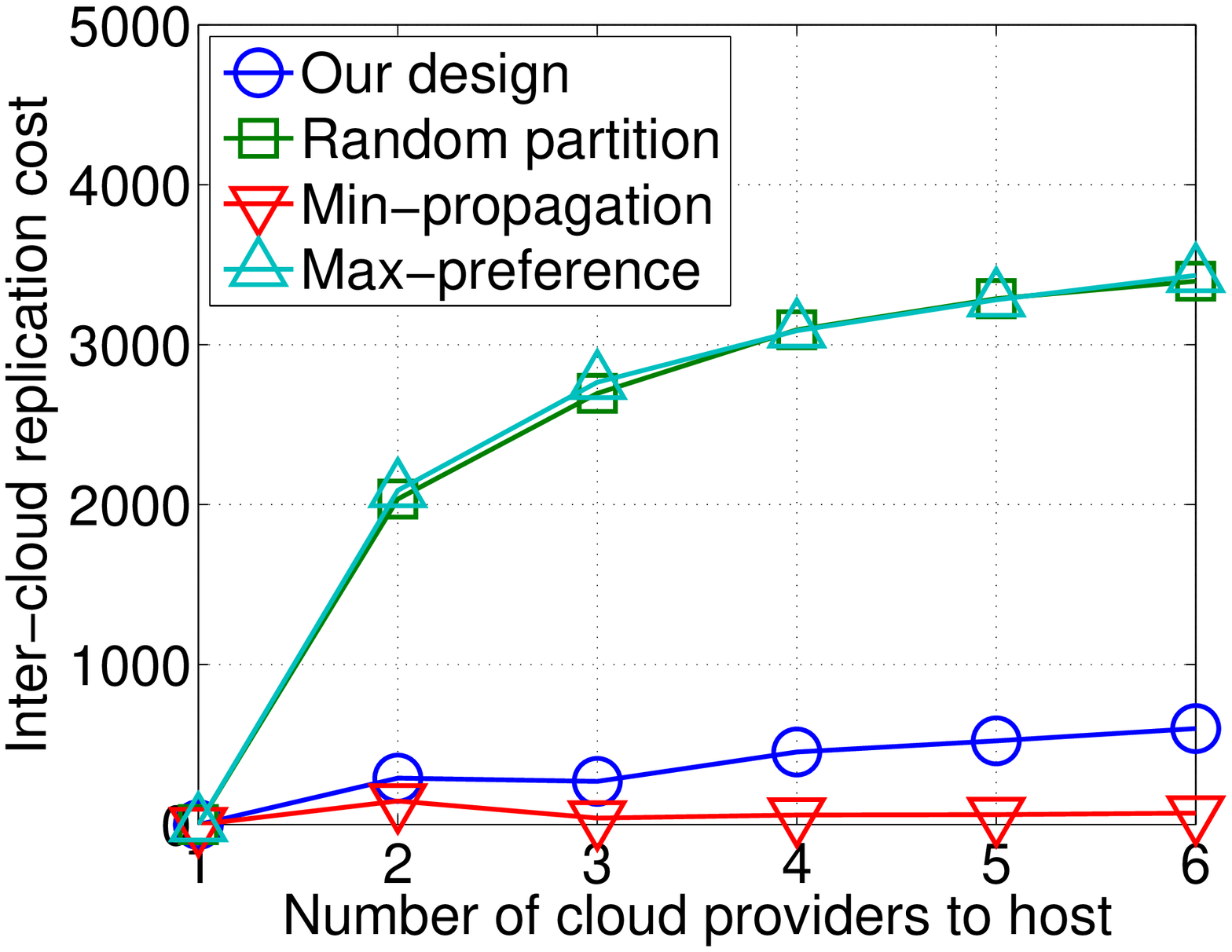}
		\centerline{\parbox[t]{\linewidth}{\scriptsize (b) Replication cost versus the number of cloud providers to host the instant social video sharing system.}}
	\end{minipage}     
	\caption{Comparison of different multi-cloud hosting algorithms under different number of cloud providers.}
	\label{fig:eval-inter-cloud}
\end{figure}

We study their performance by varying the number of cloud providers that are available for the instant social video sharing system to perform the multi-cloud hosting. In Fig.~\ref{fig:eval-inter-cloud}(a), we first study the satisfaction of users' cloud-provider preference. We observe that the user preference increases in both our design and the max-preference strategy; while the other two strategies cannot benefit from the availability of more cloud providers. We also observe that the max-preference strategy outperforms our design by about $12\%$ when all the $6$ cloud providers can be utilized in the multi-cloud hosting.

However, the max-preference algorithm incurs much larger inter-cloud replication cost due to the social propagation over the social connections between users hosted with different cloud providers. The curves in Fig.~\ref{fig:eval-inter-cloud}(b) illustrate the propagation cost against the number of available cloud providers. We observe that both our design and the min-propagation strategy remain very low inter-cloud propagation cost as the number of cloud providers increases, while the cost increases in both the max-preference and random strategies as the number of cloud providers increases. The inter-cloud propagation cost in the max-preference strategy is about $6$ times larger than that in our design when all the cloud providers can be utilized. The results indicate that our design can well balance users' cloud-provider preference and the cost of inter-cloud propagation.

\subsubsection{Effectiveness of the Heuristic Algorithm}

Since finding the optimal solution for deploying the social video sharing system is generally NP-Hard, we present the effectiveness of our heuristic algorithm. In particular, we compare the combined objective defined in Eq.~\ref{eq:obj} achieved by our algorithm, with the optimal value achieved by a brute-force searching. In this experiment, we generate a graph with $10$ nodes (users), who have random preferences of $3$ cloud providers.

By varying the number of the directed social connections between them, we compare the combined objective value achieved by both algorithms in Fig.~\ref{fig:optimal-cmp}. We observe that when the number of social connections (edges) is under $30$ (\emph{i.e.}, $1/3$ of all possible social connections), our algorithm achieves similar performance as the optimal solution does --- in a real social graph, the number of social connections is much smaller than that \cite{mislove2007measurement}. Thus, our algorithm has relatively good performance for the real online social network systems.

\section{Related Work} \label{sec:relatedwork}

\subsection{Growth of Online Social Media and Cloud-based Hosting}

The philosophy of social media is to let users in an online social network not only generate the content, but also disseminate the generated content through social connections \cite{kaplan2010users}, including the ``following'' relationship in a Twitter-like system, and the ``friending'' relationship in a Facebook-like system.

Cloud computing has been widely used to deploy the social media systems. As both the number and the geographic distribution of the users in an online social media service are expanding, hosting such a social media system can take full advantage of the elastic cloud resource \cite{pujol2010little}. In \cite{pujol2010little}, social graph is studied for locality partition, while in our study, we design partition by studying social propagation, which is determined by not only the social graph, but also user behaviors. 

There are several works on hosting a social media system with different cloud computing platforms. Cheng \emph{et al.}~\cite{cheng2011load} have studied the migration of socialized videos in YouTube to the cloud so as to balance the load between servers. Wu \emph{et al.}~\cite{wuscaling2012} have studied how to scale the social media service using the geo-distributed cloud resource. In our previous study, we presented that using a edge cloud framework, users can benefit from downloading from local servers \cite{zhi-tomccap-psar2013} in social video streaming.

\BeginRevision

\subsection{Instant Social Video Delivery}

Given the crowdsourced content capturing and sharing, the preferred length of online videos becomes shorter and shorter. Vine and Weishi are representative services that enable users to create ultra-short video clips, and instantly post and share them with their friends. Taking Vine as a case study, Zhang et al.~\cite{zhang2014understand} show that the instant social videos have short lifetime and highly skewed popularity that fast decays over time. Videos in these social trending media become more fragmented and instantaneous, which have challenged today's content replication and streaming strategies.

\EndRevision

\subsection{Social Propagation}

In a social media system, contents propagate among users due to a variety of social activities, \emph{e.g.}, users can \emph{reshare} contents that are originally generated by their friends, so as to make the contents available to more people in the online social network.
To efficiently serve the social media contents, the propagation information has to be considered for the service deployment \cite{yoganarasimhan2012impact}. 

In an online social network, contents can be dynamically shared by social groups with very different size and geographic distribution \cite{backstrom2010find}. As a result, propagation inference has become an important factor for improving the performance of social media services --- a number of research efforts have been devoted to studying the content propagation in social media \cite{kwak2010twitter}, including the traditional message propagation in Twitter-like microblogging systems \cite{petrovic2011rt}, as well as the video propagation in YouTube-like systems \cite{li2012understanding}. 
Xu \emph{et al.}~\cite{xu2014forecasting} studied how to forecast video popularity of social video contents, and observed that social propagation is a critical factor for predicting social video contents.

\subsection{Graph Partition for Distributed Social Media}

A fundamental problem in hosting social video contents with a distributed system is the partition of contents and users in the social graph. Tran \emph{et al.}~\cite{tran2012s} have studied the partition of contents in an online social network by taking users' social relationship into consideration. Newman \emph{et al.}~\cite{newman2004finding} have studied the community structure in the social network. Pujol \emph{et al.}~\cite{pujol2010little} have designed a social partition and replication middleware where data from a user's friends can be co-located at load-balanced servers. Carrasco \emph{et al.}~\cite{carrasco2011partitioning} have proposed to partition the social graph by dividing users' activities into different time phases, since the propagation levels between two users vary over time. 

These works have studied the hosting of a social media system in the context of a single cloud provider, or the cloud servers are treated equally even when they are allocated from different cloud providers --- the replication roadblock across the boundary among different cloud providers does not exist. However, this assumption is no longer true under the pricing scheme of today's cloud providers \cite{amazonec2price}.

In this paper, we seek to design a multi-cloud hosting strategy based on a social graph partition, which jointly takes users' cloud-provider preference, the content propagation between users, and the replication roadblock across the boundary between cloud providers caused by their pricing schemes into account.

\section{Concluding Remarks} \label{sec:conclusion}

In this paper, we have studied hosting an instant social video sharing service with multiple cloud providers. Our measurement studies not only confirm the benefit of the multi-cloud hosting, but also reveal several guidelines for the multi-cloud hosting design. The multi-cloud hosting problem to optimize a combination of satisfying users' cloud-provider preference and reducing the cost caused by inter-cloud social propagation is proven to be NP-hard in general. We design a heuristic algorithm to solve the problem, by iteratively partitioning a propagation-weighted social graph --- based on an initial preference-aware partition, a propagation-aware re-hosting dynamically reduces the inter-cloud propagation of the most active social connections. Trace-driven simulations further demonstrate that our heuristic can efficiently solve the multi-cloud hosting problem, and our design achieves a good balance of the two objectives under acceptable complexity --- with only $12\%$ user preference degradation, our algorithm reduces $5/6$ of the inter-cloud data transfer cost.

\section*{Acknowledgment}

We thank the Tsinghua-Tencent Joint Laboratory for providing the valuable traces used in our study. This work is supported in part by the National Basic Research Program of China (973) under Grant No.~2011CB302206.


\end{document}